\documentclass[twocolumn,
showpacs,preprintnumbers,
showkeys,
amsmath,amssymb,
prl,reprint,
superscriptaddress
]{revtex4-1}

\usepackage{graphicx}
\usepackage{dcolumn}
\usepackage{bm}
\usepackage{hyperref}
\usepackage{textcomp}

\begin{document}


\title{24~\textmu m spin relaxation length in boron nitride encapsulated bilayer graphene}

\author{J. Ingla-Ayn\'es}
 \email{J.Ingla.Aynes@rug.nl}
 \affiliation{Physics of Nanodevices, Zernike Institute for Advanced Materials, University of Groningen, The Netherlands}
\author{M. H. D. Guimar\~aes}
\affiliation{Physics of Nanodevices, Zernike Institute for Advanced Materials, University of Groningen, The Netherlands}
\affiliation{Kavli Institute at Cornell, Cornell University, Ithaca, NY 14853, USA}
\author{R. J. Meijerink}
\affiliation{Physics of Nanodevices, Zernike Institute for Advanced Materials, University of Groningen, The Netherlands}
\author{P. J. Zomer}
\affiliation{Physics of Nanodevices, Zernike Institute for Advanced Materials, University of Groningen, The Netherlands}
\author{B. J. van Wees}
\affiliation{Physics of Nanodevices, Zernike Institute for Advanced Materials, University of Groningen, The Netherlands}

\date{June 1, 2015}

\begin{abstract}
We have performed spin and charge transport measurements in dual gated high mobility bilayer graphene encapsulated in hexagonal boron nitride. Our results show spin relaxation lengths $\lambda_s$ up to 13~\textmu m at room temperature with relaxation times $\tau_s$ of 2.5~ns. At 4~K, the diffusion coefficient rises up to 0.52~m$^2$/s, a value 5 times higher than the best achieved for graphene spin valves up to date. As a consequence, $\lambda_s$ rises up to 24~\textmu m with $\tau_s$ as high as 2.9~ns. We characterized 3 different samples and observed that the spin relaxation times increase with the device length. We explain our results using a model that accounts for the spin relaxation induced by the non-encapsulated outer regions.
 
\end{abstract}

\maketitle


The ability to transport spin currents over long distances is a major requirement to achieve new functionalities using spin information \cite{Fabian_Acta}. For this purpose, graphene and its multilayer forms are predicted to be ideal materials. The low spin orbit coupling provided by the low atomic mass of the carbon atoms \cite{ReviewFabian} together with their outstanding electronic properties \cite{RevModPhysGraph} make them great candidates for spin transport devices.

Even though theoretical predictions suggest spin relaxation times ($\tau_s$) up to 100~ns in single layer pristine graphene \cite{TeoriaSOC}, the first experimental results for graphene on SiO$_2$ showed $\tau_s=$ 150~ps and spin relaxation lengths ($\lambda_s$) of about 1.5~\textmu m \cite{NikoSpinGraf}, opening the debate on which relaxation mechanism rules the spin relaxation in graphene\cite{ReviewFabian, RevRocheVal}. Following the first experiments on single layer graphene, other experiments \cite{ BLGKawakami,BLGAachen} focussed on the spin transport properties of bilayer graphene \cite{ElPropOfBLG, RegenBLG, NatGapBLG}. These experiments reported spin relaxation times up to 6.2~ns at low temperatures \cite{BLGKawakami} and 2~ns at room temperature \cite{BLGAachen}.

A recent experiment using hexagonal boron nitride (BN) to encapsulate single layer graphene achieved spin relaxation lengths up to 12~\textmu m at room temperature \cite{PRLMarcos}. Another experiment, using partially suspended multilayer graphene that was covered by a BN flake, achieved room temperature $\tau_s$ up to 3.9~ns in trilayer graphene \cite{AachenNanoletters} showing the potential of graphene/BN heterostructures for spin transport.    

In this rapid communication we report spin transport in high mobility bilayer graphene (BLG). Our samples consist of BLG that is partially encapsulated between two BN flakes. The stacks are fabricated using a dry transfer technique that allows us to get very clean interfaces without extra cleaning steps \cite{1Dcontact, FastPickUp}. After this process, the (BN/BLG/BN) stack remains covered by a PC film that is removed by keeping it in chloroform at 50~$^\circ$C for 5 hours. This step is followed by an annealing at 250~$^\circ$C for 24 hours with an Ar/H$_2$ flow.
The contacts  are defined using e-beam lithography and designed with different widths (from 0.1 to 0.5~\textmu m) to ensure different coercive fields. The deposition was done using e-beam evaporation of the corresponding films. The TiO$_2$ tunnel barriers were evaporated in 2 steps of 0.4~nm of Ti, followed by 15 minutes of oxidation in pure oxygen gas.
The 65~nm thick Co contacts and the top gate were evaporated in one step followed by an Al capping layer.

\begin{figure}[h]
	\centering
		\includegraphics[width=0.5\textwidth]{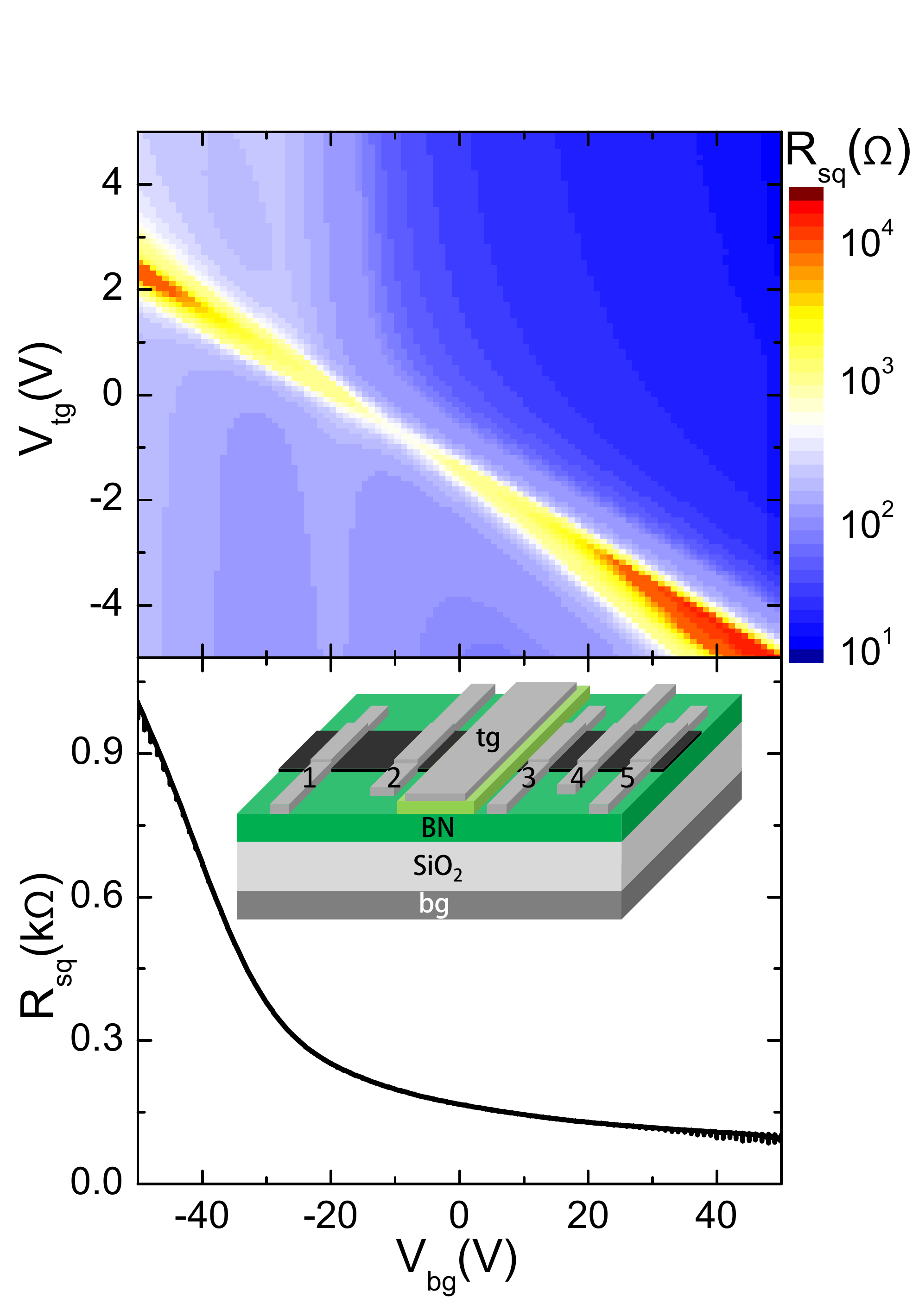}
	\caption{(a) Square resistance obtained between contects 2 and 3 (in color scale) with respect to $V_{tg}$ (y axis) and $V_{bg}$ (x axis) (b) Square resistance of the non-encapsulated region measured between contacts 3 and 4. Inset: schematics of the device.}
	\label{fig:figure-1}
\end{figure}

The sample configuration is shown in the inset of Fig. \ref{fig:figure-1}. The bilayer graphene, in black, is encapsulated between the top and bottom BN. The flake is fully encapsulated in the central region while both left and right sides are not encapsulated but only supported on a bottom BN. This configuration allows us to have ferromagnetic contacts at both sides of the sample while keeping the central region protected. This nonhomogeneous design also has consequences for the analysis of the data as we explain below.

The top-gate together with the Si back-gate (Fig. \ref{fig:figure-1} (b) inset), allow us to simultaneously control the electric field $\overline{E}=\epsilon_{bg}(V_{bg}-V_{bg}^{(0)})/2d_{bg}-\epsilon_{tg}(V_{tg}-V_{tg}^{(0)})/2d_{tg}$ and the carrier density $n=\epsilon_0 \epsilon_{bg}(V_{bg}-V_{bg}^{(0)})/ed_{bg}+\epsilon_0 \epsilon_{tg}(V_{tg}-V_{tg}^{(0)})/ed_{tg}$ applied to the dual-gated region. Here $e$ is the electron charge, $\epsilon_0$ is the vacuum permittivity, $\epsilon_{bg(tg)}\approx 3.9$ the dielectric permittivity, $d_{bg(tg)}$ the thickness of the dielectric, $V_{bg(tg)}$ the applied gate voltage and $V_{bg(tg)}^{(0)}$ the voltage at charge neutrality point of the back-gate (top-gate) respectively\cite{NatGapBLG}.

We have characterized 3 devices (A, B and C) showing similar results at room temperature and 4~K. The results are shown in Table \ref{table1}. From there we can see that $\tau_s$ seems to depend on the length of the encapsulated region and, even though device C shows a higher  spin diffusion coefficient ($D_s$) than device B, indicating better electronic quality, $\tau_s$ of device B is more than 2 times longer than the one of device C. This nonstraightforward connection between electronic quality and $\tau_s$ seems to be in agreement with the results shown in \cite{KawakamiNanoletters} for single layer graphene, while the length dependence can be explained by the effect of the invasive contacts and the lower quality of the nonencapsulated regions being reduced increasing the contact separation \cite{ContactsHanle, SarojCVD}. As a consequence, we believe that even longer spin relaxation times can be achieved by making longer devices.
\begin{table}[t]
    \caption{Spin parameters obtained at the gate combination giving the longest $\lambda_s$ for 3 devices with different length of the encapsulated regions. $L_{enc}$ is the length of the encapsulated region and $d_{2-3}$ is the separation between the inner contacts}
        \begin{ruledtabular}
        \begin{tabular}{c c c c c c c}
            \renewcommand{\arraystretch}{2}
            Dev. & $L_{enc}$ (\textmu m)& $d_{2-3}$ (\textmu m)& T (K)& $\tau_s$(ns) & $D_s$(m$^2$/s) & $\lambda_s$(\textmu m) \\
            \hline
             A & 13.2& 14.6 & 300 & 2.5 & 0.07 & 13\\
             & & &4 & 2.9 & 0.2 & 24\\
             B & 8.5 &10.3& 300 & 1.1 & 0.03 & 5.7\\
             & & &4 & 1.9 & 0.05 & 9.7\\
             C & 6.2 &7.8& 300 & 0.32 & 0.04 & 3.6\\
             & & &4 & 0.45 & 0.07 & 5.6\\
    \end{tabular}
    \end{ruledtabular}
    \label{table1}
\end{table}
 


From now on we will discuss the results obtained at 4~K for device A.
The contact resistances range between 280~$\Omega$ and 2.7~k$\Omega$. These low values are a consequence of imperfect tunnel barriers and affect the spin transport measurements \cite{ContactsHanle}.
 
To characterize the charge transport properties of our device we carried out standard 4 probe measurements. We determine the resistances of the encapsulated and non-encapsulated regions by measuring the voltages ($V_{2-3}$) between contacts 2 and 3 and ($V_{3-4}$) between 3 and 4 while driving a current ($I_{1-5}$) between 1 and 5 (Fig. \ref{fig:figure-1} (b)). In Fig. \ref{fig:figure-1} (a) we show the square resistance ($R_{sq}$) of the encapsulated region as a function of the back-gate voltage $(V_{bg})$ and the top-gate voltage $(V_{tg})$. The charge neutrality point appears as a line with a slope $-C_{tg}/C_{bg}= -0.078$ showing a resistance minimum at $V_{bg}= -$8~V, $V_{tg}= -$0.7~V. Taking into account that this point has zero carrier density and zero electric field we can estimate the doping at the top and bottom sides of the flake: $n_{bg}^{(0)} \approx n_{tg}^{(0)} \approx$5.5$\times 10^{15}$~m$^{-2}$. 

The resistance increases at both sides of the charge neutrality line, reaching up to 38~k$\Omega$ at an electric field of 0.69~V/nm. This is caused by the opening of a gap driven by the electric field \cite{NatGapBLG}. One can also distinguish two $V_{tg}$ independent features coming from the non-top-gated region between the inner contacts. One comes from the sides of the encapsulated regions that are non-top-gated and show a charge neutrality point around $V_{bg}= -$19~V. The other comes from the nonencapsulated regions and its square resistance is shown in Fig. \ref{fig:figure-1} (b).  

The mobility ($\mu$) obtained for this sample at $V_{bg}= -8$~V (zero electric field) is 16~m$^2$/Vs at the electron side. The value is obtained using the formula $R_{sq}=1/n e \mu+\rho_s$ where $\rho_s \approx 57$~$\Omega$ is the $n$-independent resistivity component coming from the effect of the resistance of the non-top-gated regions between the inner contacts and short range scattering \cite{MobGeimPRL}. We have also confirmed that the resistance of the outer regions of the sample does not depend on $V_{tg}$. 

In Fig. \ref{fig:figure-1}(b) we show the $V_{bg}$ dependence of one of the outer regions' resistance. As it can be seen from the graph, the charge neutrality point is below $V_{bg}= -$50~V and falls outside our gate range. We attribute this to the contamination given by the polymers used during fabrication.

\begin{figure}[h]
	\centering
		\includegraphics[width=0.5\textwidth]{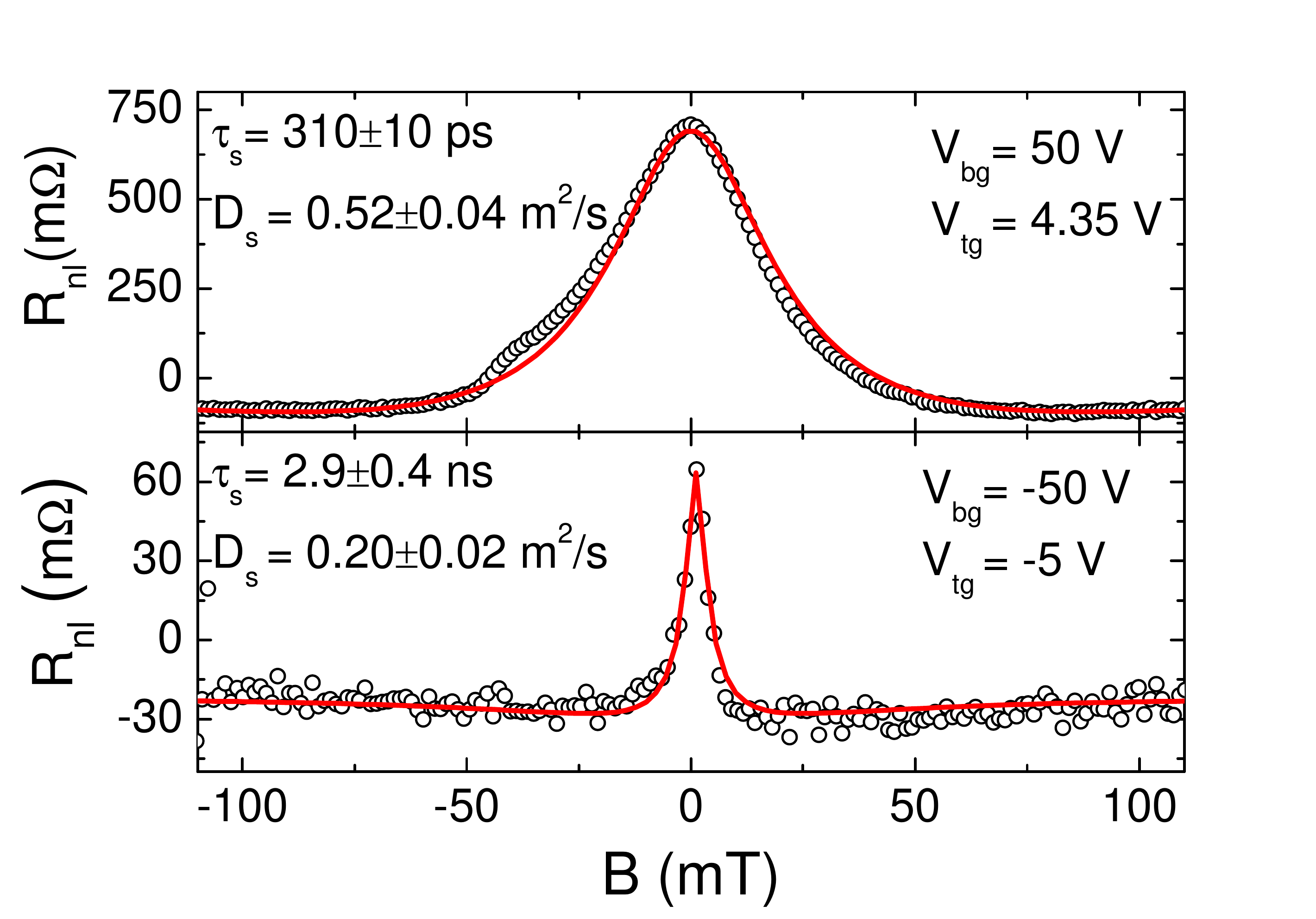}
	\caption{Hanle precession curves obtained at $V_{bg}= -$50~V, $V_{tg}= -$5~V (top panel) and $V_{bg}= $50~V, $V_{tg}= $4.35~V (bottom panel) with the corresponding fitting curves and extracted spin parameters.}
	\label{fig:figure-2}
\end{figure}
To measure the spin transport properties of the encapsulated region, we used the standard non-local geometry \cite{John_Sils} sending a current ($I_{1-2}$) between contacts 1 and 2 and detecting the voltage ($V_{3-5}$) between 3 and 5. When applying an out-of-plane magnetic field the spins undergo Larmor precession. Measuring $R_{nl}=V_{3-5}/I_{1-2}$ while sweeping the magnetic field we obtained the so called Hanle precession curves. 
To eliminate spin-independependent effects we have taken Hanle curves for parallel and antiparallel magnetic configurations of the inner contacts and substracted them $R_{nl}=(R_{nl}^{par}-R_{nl}^{anti})/2$ where $R_{nl}^{par(anti)}$ is the nonlocal resistance in the parallel (antiparallel) magnetic configuration \cite{ContactsHanle}. The magnetizations of the contacts are tuned applying an in-plane magnetic field.

 In Fig. \ref{fig:figure-2} we show two Hanle curves taken at $V_{bg}= 50$~V, $V_{tg}= 4.35$~V and $V_{bg}= -50$~V, $V_{tg}= -5$~V, corresponding to the top right and bottom left corners in the color plot of Fig \ref{fig:figure-1} (a). The spin relaxation time and diffusion coefficients are extracted from these curves by fitting them with the solution of the Bloch equations \cite{Fabian_Acta, John_Sils} including a small offset \cite{HanleOffset}. 
 
 The spin signal at $V_{bg}= 50$~V, $V_{tg}= 4.35$~V (Fig. \ref{fig:figure-2} top panel) is 10 times higher than the one at $V_{bg}= -50$~V, $V_{tg}= 5$~V (Fig. \ref{fig:figure-2} bottom panel). This is most likely due to the low resistance of our contacts. When the contact resistance is in the order of the spin resistance of the channel ($R_{\lambda}=R_{sq}\lambda_s/W$, here $W$ is the width of the sample), part of the injected spin accumulation relaxes back to the contacts instead of diffusing in the channel, reducing the effective injection efficiency \cite{ContactsHanle, MgOAachen} from 12\% to 2\%. This effect is ruled by the resistance of the outer regions (where the contacts are placed) and it is $V_{tg}$ independent.
 
\begin{figure}[h]
	\centering
		\includegraphics[width=0.5\textwidth]{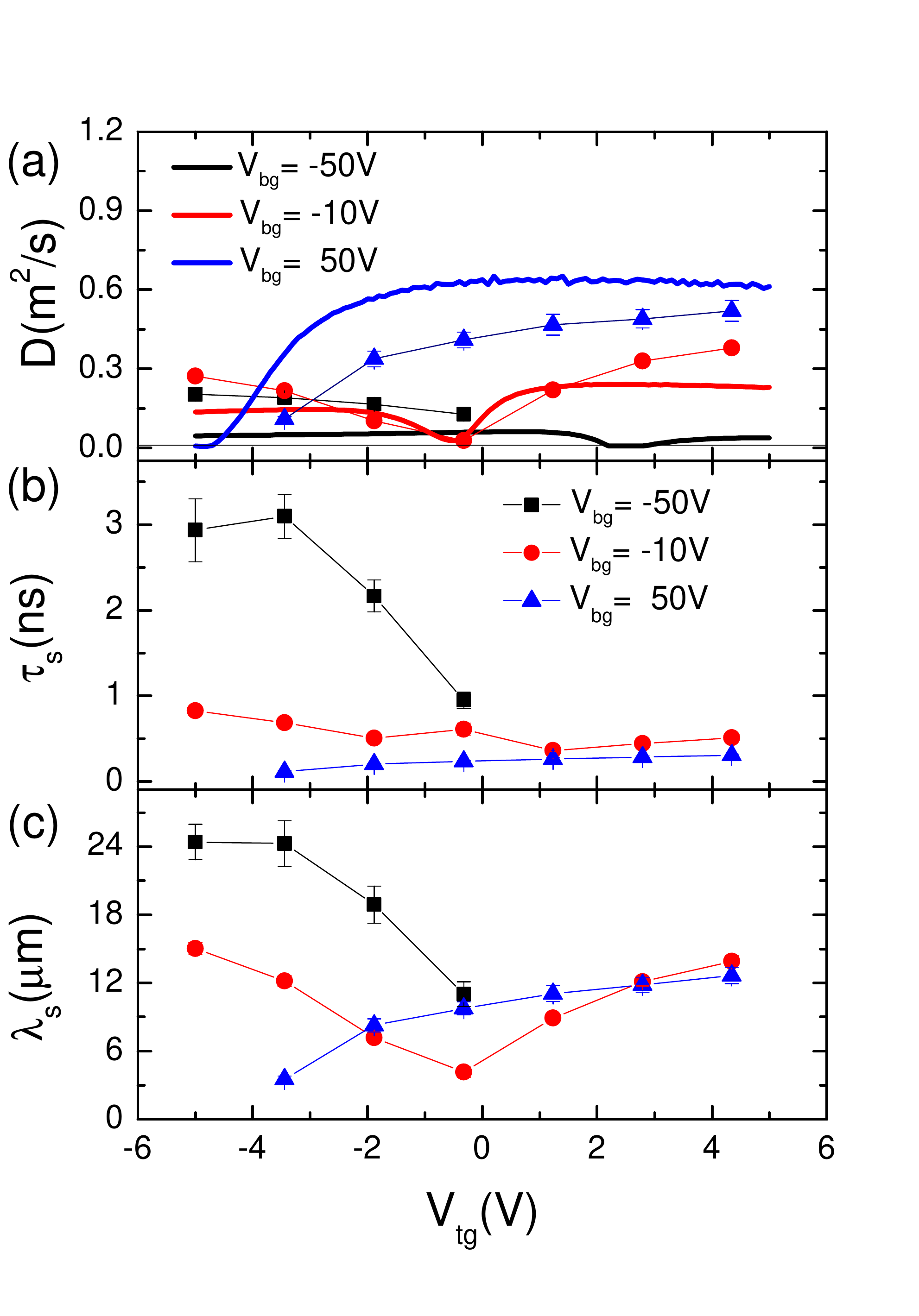}
	\caption{Spin transport parameters obtained by fitting the Hanle curves at 4~K as a function of $V_{tg}$ for $V_{bg}=$ $-$50,$-$10 and 50~V. (a) Spin diffusion coefficient (dots) compared with the charge diffusion coefficient (solid lines), (b) spin relaxation time and, (c) the relaxation length.  The lines connecting the spin parameters are a guide to the eye.}
	\label{fig:figure-3}
\end{figure}
In Fig. \ref{fig:figure-3} (a) we show the spin and charge diffusion coefficients  for 3 different backgate voltages in dots and solid lines respectively. The charge diffusion coefficients ($D_c$) were obtained using the Einstein relation $1/D_c=e^2 R_{sq} \nu (E_F)$, where $\nu (E_F)$ is the density of states at the Fermi energy and $e$ the electron charge. $R_{sq}$ was taken from Fig. \ref{fig:figure-1} (a) and corrected by substracting the resistance of the nonencapsulated regions between the inner contacts. 

At $V_{bg}= -50$ V one can observe a substantial difference between $D_c$ and $D_s$. We attribute this to the fact that the gate induced doping of the encapsulated and nonencapsulated regions have different signs, creating $pn$ junctions of unknown widths at the boundaries. Since these boundaries are between the inner contacts, the measured square resistances are no longer characteristic of the channel itself but of the junctions. This affects the determination of $D_c$.

 At $V_{bg}= 50$~V both encapsulated and non-encapsulated regions are electron-doped and $D_c$ shows better agreement with respect to $D_s$, supporting the validity of the parameters obtained from the Hanle measurements. The minor discrepancy in this case can be attributed to the small resistance of the encapsulated non-top-gated region that was not substracted from the calculation of $R_{sq}$. $D_c$ and $D_s$ reach values above 0.5~m$^2$/s, 5 times higher than the best achieved for BN encapsulated single layer graphene spin valve devices \cite{PRLMarcos}.

At $V_{bg}= -10$~V (approximately zero electric field) we see that close to the charge neutrality point ($V_{tg} \approx 0$~V) there is a better agreement between $D_c$ and $D_s$ than at high carrier densities. This can be explained taking into account that close to the charge neutrality point the square resistance of the encapsulated region is high enough to dominate the measurement of $R_{sq}$, but at high carrier densities, the square resistance of the encapsulated region is small and the contributions of the non-top-gated regions become relevant. 

 In Fig. \ref{fig:figure-3} (b) there is a strong dependence of the spin relaxation time on $V_{bg}$. At $V_{bg}= -$50~V, the relaxation time reaches 3~ns while at 50~V a maximum of $\tau_s=  $310~ps is obtained. This reduction of a factor 10 in $\tau_s$ is in agreement with the results in \cite{PRLMarcos} and can be explained as an effect of the change in $R_\lambda$ of the nonencapsulated regions. As the spin resistance of these regions increases, their influence on the spin relaxation is reduced. This effect occurs because the spins relax predominantly at the regions with the lower $R_{\lambda}$.
 
  The opposite effect occurs when opening a gap in the encapsulated region and its square resistance increases with respect to the one of the non-encapsulated part. Since $\tau_s$ is longer in the encapsulated region than in the outer ones, $R_{\lambda}$ gets orders of magnitude higher than the one of the outer part. As a consequence, the spin relaxation is dominated by the non-encapsulated regions and the amplitude of the spin signal vanishes. This effect explains why we could not measure spin signals at $V_{bg}= -50$~V and positive $V_{tg}$.
  
  In Fig. \ref{fig:figure-3} (c) we show the spin relaxation lengths calculated using the formula $\lambda_s=\sqrt{D_s\tau_s}$. Apart from its carrier density dependence, we can see that $\lambda_s$ goes up to 24~\textmu m, the highest value achieved up to date by fitting Hanle measurements in nonlocal geometry. 
  Note that, even though spin relaxation lengths up to 280~\textmu m were estimated from local 2 probe measurements at 4~K for epitaxial graphene on a SiC substrate \cite{Fert100um}, no Hanle measurements were done to support these values.

%
\begin{figure}[h]
	\centering
		\includegraphics[width=0.4\textwidth]{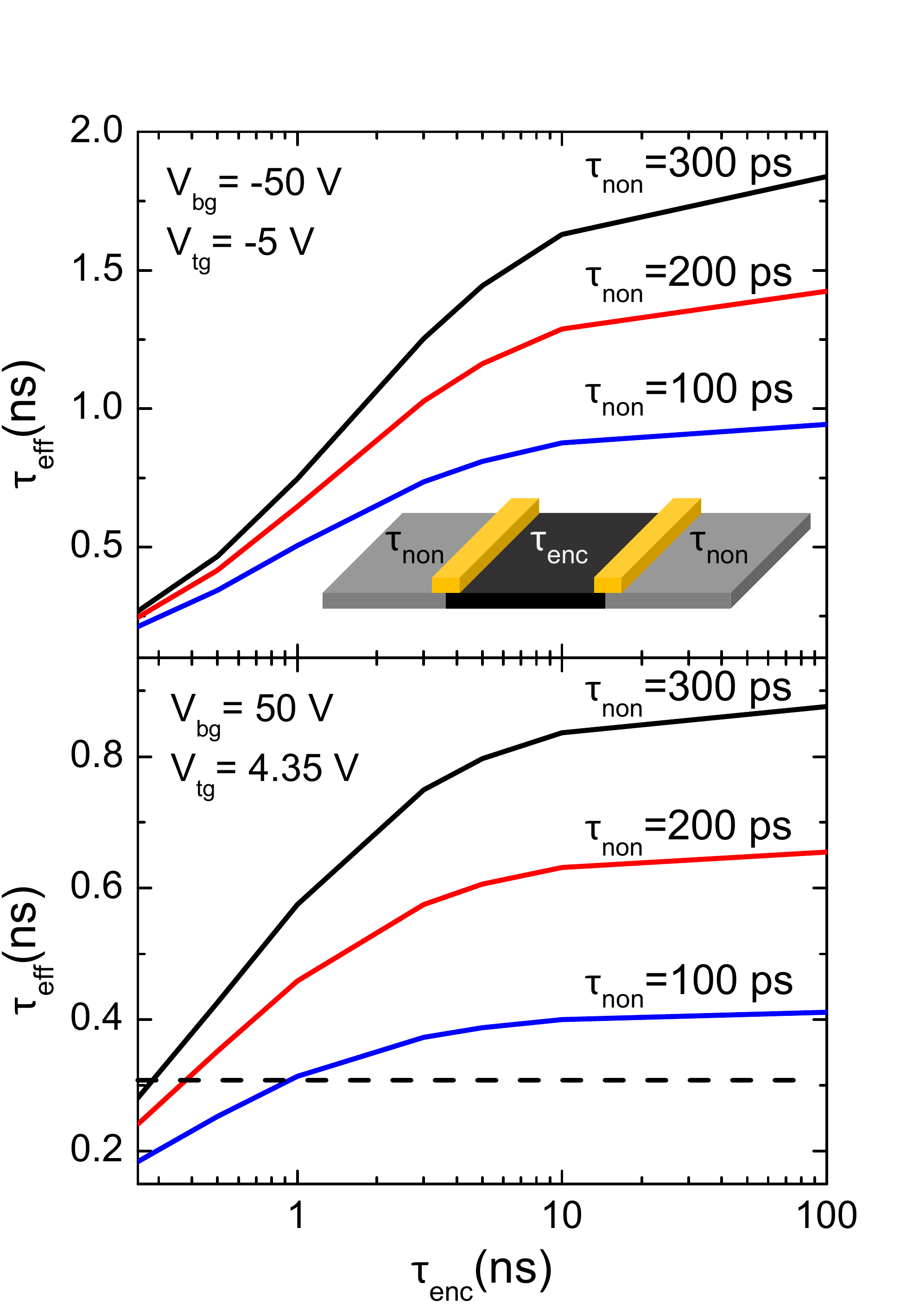}
	\caption{Effective spin relaxation time in the system as a function of $\tau_{enc}$ for 3 different values of $\tau_{non}$ at $V_{bg}= -50$~V and  $V_{tg}= -5$~V (a) and  $V_{bg}= 50$~V and  $V_{tg}= 4.35$~V (b), the dashed line is the experimental value of $\tau_{eff}$, taken from Fig. \ref{fig:figure-2} (top panel). The inset shows a cartoon device of the simulated system.}
	\label{fig:figure-4}
\end{figure}
Since the spins probe the whole device (inner and outer regions), we have to account for the nonhomogeneity of our sample to explain our results \cite{PRLMarcos}. For this reason, we use the same model as in \cite{PRLMarcos} and \cite{MarcosSuspended}, where we solve the Bloch equations for a nonhomogeneous system consisting of a central region sandwiched by 2 regions as shown in the inset of Fig. \ref{fig:figure-4}. We set the spin and charge transport parameters ($\tau_{s}$, $D_{s}$ and $R_{sq}$) for the 3 regions assuming that the outer regions are identical. After simulating the corresponding Hanle curves, we fit them using a homogeneous model as we have done with our experimental data. From this procedure, we obtain the effective relaxation time of the system ($\tau_{eff}$). 

The effect of invasive contacts is taken into account using the spin transport parameters obtained from Hanle measurements carried out at the non-encapsulated regions. Since the contact separation in these regions is between 1 and 2~\textmu m, the extracted parameters are strongly affected by the low contact resistances \cite{ContactsHanle}.

In Fig. \ref{fig:figure-4} we plot $\tau_{eff}$ as a function of the spin relaxation time in the encapsulated region ($\tau_{enc}$) for 3 different values for the spin relaxation time in the non-encapsulated region ($\tau_{non}$). The resistance values used for the central region are the ones used to calculate $D_c$ in Fig. \ref{fig:figure-2} (a). For the diffusion coefficient in the encapsulated region ($D_{enc}$) we have used the values of $D_s$ extracted from the experiments at the encapsulated region. This is justified since $D_{s}$ is not affected by the outer regions\cite{MarcosSuspended, PRLMarcos}. The square resistance of the non-encapsulated region is taken from Fig. \ref{fig:figure-1} (b) and the diffusion coefficient ($D_{non}$) is taken from the experimental Hanle curves obtained at the outer region.

In Fig. \ref{fig:figure-4} (a), where $V_{bg}= -50$~V and $V_{tg}= -5$~V, the maximum $\tau_{eff}$ obtained from the simulations reaches 1.8~ns for $\tau_{enc}= 100$~ns and $\tau_{non}= 300$~ps. This value is still below the 2.9~ns obtained from the fittings of the encapsulated data. This discrepancy can be explained taking into account that, as discussed above, there are $pn$ junctions in the boundaries between encapsulated and non-encapsulated regions and this is not taken into account in our simulations.

The simulations at $V_{bg}= 50$~V and $V_{tg}= 4.35$~V are shown in Fig. \ref{fig:figure-4} (b). The dashed line at 310~ps corresponds to the value obtained from the fittings of the experimental results at the encapsulated region. The intersections between the simulated curves and the dashed line give us the possible values of $\tau_{enc}$. From the fittings of the Hanle curves measured at the outside regions, we obtained $\tau_{non}\approx$100~ps. As a consequence, from our simulations, $\tau_{enc}\approx 1$~ns. Using this relaxation time and $D_{enc}= 0.52$~m$^2$/s, we obtain a spin relaxation length of 22~\textmu m, close to the 24~\textmu m at $V_{bg}= -50$~V and $V_{tg}= -5$~V.

In conclusion, we have characterized 3 boron nitride encapsulated bilayer graphene devices with 13.2, 8.5 and 6.2~\textmu m long encapsulated regions showing consistent behavior. 

Our results show that the measured $\tau_s$ depends on the length of the encapsulated region in agreement with \cite{PRLMarcos}.

The results obtained for the longer device show unprecedented large spin diffusion coefficients up to 0.52~m$^2$/s at 4~K, 5 times higher than the best achieved for single layer graphene using the same geometry \cite{PRLMarcos}. As a consequence, the spin relaxation length rises up to 13~\textmu m at room temperature and 24~\textmu m at 4~K.
  
Our simulations using a three regions model show that the measured spin relaxation times of 2.5~ns at room temperature and 2.9~ns at 4~K are most likely limited by the outer regions, suggesting that it is possible to transport spin information over even larger distances in the used geometry by increasing the length of the encapsulated region. According to this result, higher spin relaxation times can be achieved by making longer devices \cite{SarojCVD}.

We would like to acknowledge J. C. Brant, R. Ruiter and J. C. Leutenantsmeyer for useful discussions and J. G. Holstein, H. M. de Roosz, H. Adema and T. Schouten for technical support.
The research leading to these results has received funding from the People Programme (Marie Curie Actions) of the European Union's Seventh Framework Programme FP7/2007-2013/ under REA grant agreement n$^{\circ}$607904-13 Spinograph, the Dutch Foundation for Fundamental Research on Matter (FOM), the European Union Seventh Framework Programme under grant agreement n$^{\circ}$604391 Graphene Flagship, the Netherlands Organization for Scientific Research (NWO), NanoNed and the Zernike Institute for Advanced Materials.
\bibliography{rapcombib}
\end{document}